\documentclass[aps,prb,twocolumn,groupedaddress,amsmath,letterpaper]{revtex4}

\usepackage{graphicx}

\begin{document}

\title{Scattering matrix of the boundary of a nonlocal metamaterial}

\author{Chris Fietz$^{1}$}
\email[Email: ]{fietz.chris@gmail.com}
\author{Costas M. Soukoulis$^{1,2}$}
\affiliation{$^1$Ames Laboratory and Department of Physics and Astronomy, Iowa State University, Ames, Iowa 50011, USA \\ $^2$Institute of Electronic Structure and Laser, FORTH, 71110 Heraklion, Crete, Greece}

\begin{abstract}
We present a simple model of the interface between a local homogeneous medium and a potentially nonlocal metamaterial/photonic crystal.  This model allows us to calculate the scattering matrix elements of the interface for a plane wave of light normally incident upon the interface from either direction.  The resulting scattering matrix provides insight into the non-Maxwellian boundary conditions present at the interface between a homogeneous medium and a metamaterial/photonic crystal with strong spatial dispersion.  We present the model mathematically.  As an example, the model is used to calculate the scattering matrix of the interface between vacuum and a simple photonic crystal.  Several tests of the calculated scattering matrix elements are presented.  Finally, we used the results of the scattering model to postulate possible forms for the non-Maxwellian boundary conditions.
\end{abstract}


\maketitle


\section{Introduction}\label{Intro}
Most of the theoretical/numerical research into metamaterials since the inception of the field~\cite{Pendry_99} has been focused on designing and modelling new types of metamaterial crystals and inclusions, as well as developing applications for metamaterials.  A much smaller subset of research has investigated different methods of characterizing metamaterials using numerical simulations, a small number of notable examples being Refs.~\onlinecite{Soukoulis_02,Li_09,Kuester_03} which characterize metamaterials (and metasurfaces) with scattering simulations as well as Refs.~\onlinecite{Smith_06,Silveirinha_07,Li_07,Simovski_07_Bloch,Fietz_10b,Alu_11,Pors_11} which use non-scattering methods to attempt to homogenize metamaterials.  Another small subset of metamaterial research, which is relevant to many of the metamaterial characterization/homogenization methods, is the investigation of boundary conditions at the interface between metamaterials and other media.  This research can be broadly divided into two categories.  First, there are \emph{additional boundary conditions}\cite{Nefedov_05,Silveirinha_06,Nefedov_06,Silveirinha_08a,Silveirinha_08b,Maslovski_10,Yakovlev_11}, which are extra boundary conditions in addition to the standard Maxwellian boundary conditions that are necessary at the interface of a medium because of the presence of multiple propagating modes due to spatial dispersion. Second, there are \emph{non-Maxwellian boundary conditions}~\cite{Simovski_07,Vinogradov_11,Kim_11}, which are modifications of the standard Maxwellian boundary conditions that become necessary in media with spatial dispersion, even in the absence of extra propagating modes.

In this paper we only consider media supporting single propagating modes and as such are only concerned with non-Maxwellian boundary conditions.  In particular, we ask the question: what is the scattering matrix for the interface between a homogeneous medium and a homogenized metamaterial?  To see why this is an interesting question, consider the example of the interface between two different homogeneous media and then contrast that with the metamaterial question.

Assume we have two semi-infinite homogeneous media, connected by a sharp interface, both described by isotropic, local (non-spatially dispersive), but potentially temporally dispersive constitutive parameters.  The two media, which will be labelled medium $1$ and medium $2$, are described by the scalar permittivities $\epsilon_1$ and $\epsilon_2$, as well as the scalar permeabilities $\mu_1$ and $\mu_2$, respectively.  For plane waves normally incident upon the interface, the scattering at the interface depends on the relative impedances of the two media, defined as $z_1=\sqrt{\mu_1/\epsilon_1}$ and $z_2=\sqrt{\mu_2/\epsilon_2}$.  The scattering matrix elements themselves will be labelled as $r_{ij}$ for the electric field reflection amplitude of a plane wave incident upon the interface from medium $i$, and $t_{ij}$ for the electric field transmission amplitude of a plane wave incident upon the interface from medium $i$ and transmitted into medium $j$.  Here and for the rest of the paper, all fields are assumed to be monochromatic.  For our example of two homogeneous media, the formulas for the for scattering matrix elements are 

\begin{equation}\label{refl_1}
\begin{array}{rl}
r_{12}= & \displaystyle\frac{z_2-z_1}{z_2+z_1}, \\ \\
r_{21}= & \displaystyle\frac{z_1-z_2}{z_2+z_1}, \\ \\
t_{12}= & \displaystyle\frac{2 z_2}{z_2+z_1}, \\ \\
t_{21}= & \displaystyle\frac{2 z_1}{z_2+z_1}. \\ \\
\end{array}
\end{equation}

\noindent Another convenient way to represent these scattering matrix elements, is to define the last three in terms of $r_{12}$

\begin{equation}\label{refl_2}
\begin{array}{rl}
r_{21}= & -r_{12}, \\ \\
t_{12}= & 1+r_{12}, \\ \\
t_{21}= & 1+r_{21}=1-r_{12}. \\ \\
\end{array}
\end{equation}

\noindent Eqs.~(\ref{refl_1}-\ref{refl_2}) are both results of the so called Maxwellian boundary conditions, which for normally incident waves is the requirement that the tangential electric ($\textbf{E}$) and magnetic ($\textbf{H}$) fields be continuous across a boundary.  For a metamaterial with strong spatial dispersion (or any material with strong spatial dispersion), we do not expect these boundary conditions to hold true.  This is easy to see by considering the Poynting flux in a material with strong spatial dispersion.  In such a material, a second term must be added the the standard Poynting flux~\cite{Agranovich_Energy,Landau_Energy,Kamenetskii_96}.  The total time averaged Poynting flux is 

\begin{equation}\label{Poynt_1}
\textbf{S}=\textbf{S}^0+\textbf{S}^1,
\end{equation}

\noindent where $\textbf{S}^0$ is the standard expression for the time averaged Poynting flux

\begin{equation}\label{Poynt_2}
\textbf{S}^0 = \displaystyle\frac{\mathrm{Re}(\textbf{E}\times\textbf{H}^*)}{2},
\end{equation}

\noindent and $\textbf{S}^1$ is an extra term due to spatial dispersion

\begin{equation}\label{Poynt_3}
\begin{array}{rl}
\textbf{S}^1= & -\hat{\textbf{e}}_i\displaystyle\frac{\omega}{4}\mathrm{Re}\left(\textbf{E}^{\dagger}\cdot\frac{\partial \hat{\epsilon}}{\partial k_i}\cdot\textbf{E} + \textbf{E}^{\dagger}\cdot\frac{\partial \hat{\xi}}{\partial k_i}\cdot\textbf{H} \right. \\ \\
& \left. + \displaystyle\textbf{H}^{\dagger}\cdot\frac{\partial \hat{\zeta}}{\partial k_i}\cdot\textbf{E} + \textbf{H}^{\dagger}\cdot\frac{\partial \hat{\mu}}{\partial k_i}\cdot\textbf{H} \right),
\end{array}
\end{equation}

\noindent where $\hat{\textbf{e}}_i$ is a unit vector in the $i$-th direction and there is an implicit sum over all values of $i$.  Also, $\hat{\epsilon}$ and $\hat{\mu}$ are the permittivity and permeability respectively and $\hat{\xi}$ and $\hat{\zeta}$ are the bianisotropic constitutive parameters~\cite{Kong_86}.  Here and throughout the rest of this paper we are using Heaviside-Lorentz units~\cite{Jackson_Units}, which are like Gaussian units with the $4\pi$ factor absorbed into the definition of the
charges and currents.

For a plane wave, normally incident upon an interface between a homogeneous medium and a metamaterial crystal, it is easy to see from Eqs~(\ref{Poynt_1})-(\ref{Poynt_3}) that when strong spatial dispersion is present, the tangential electric and magnetic fields cannot both be continuous.  This is because continuous tangential electric and magnetic fields would force the component of $\textbf{S}^0$ normal to the interface to be continuous.  If $\textbf{S}^1$ is non-zero due to the presence of spatial dispersion, this would prevent the total time averaged Poynting flux $\textbf{S}$ from being continuous across the interface.  Since the relations between the scattering matrix elements of the interface in Eq.~(\ref{refl_2}) are derived using the Maxwellian boundary conditions, we can expect that in the presence of strong spatial dispersion some or all of the relations in Eq.~(\ref{refl_2}) should fail.  Understanding the boundary conditions of metamaterials with spatial dispersion is essential to a complete understanding of metamaterials.  Calculating the scattering matrix of a metamaterial interface is an important first step towards understanding these new boundary conditions.  The purpose of this paper is to present a simple model for numerically calculating these scattering matrix elements.

The authors are only aware of one previous attempt to calculate the scattering matrix of a metamaterial interface~\cite{Kim_11}.  This method involves inferring the interface scattering matrix elements as well as the index of refraction of a metamaterial from the total reflection and transmission amplitudes for two different slabs of the same metamaterial with different thicknesses.  While the authors of Ref.~\onlinecite{Kim_11} acknowledge that the Maxwellian boundary conditions can fail at a metamaterial boundary, they make the assumption that the two transmission coefficients of the metamaterial interface, labelled $t_{12}$ and $t_{21}$ in this paper, are equal due to Lorentz reciprocity.  This leaves only three scattering matrix elements plus the index of refraction to be algebraically inferred from the four total scattering matrix elements of the two slabs of differing thickness.  We disagree with this assumption and point out that it is not true even for the interface between vacuum and a simple dielectric, as can be seen from Eq.~(\ref{refl_1}).

We also note that it is well known that one can use Rigorous Coupled Wave Analysis~\cite{Moharam_81} (also known as Fourier Modal Method) to calculate the scattering matrix between plane waves in a vacuum (or some other homogeneous local medium) and the Bloch modes of a metamaterial/photonic crystal, separated by a sharp interface.  While this method can correctly calculate the reflection coefficient back into vacuum, labelled $r_{12}$ in this paper, the remaining scattering amplitudes (plane wave to Bloch mode, Bloch mode to plane wave, and Bloch mode to Bloch mode) must not be confused with the scattering matrix elements calculated in this paper.  The scattering matrix elements in this paper are the ratios of electric field amplitudes of planes waves, both in the homogeneous medium and in the homogenized metamaterial/photonic crystal.

In Sec.~\ref{Sec_2} we present the numerical model for calculating the scattering matrix of the metamaterial interface.  This model is based on a similar model used for metamaterial homogenization~\cite{Fietz_10b}.  In addition, this scattering model relies upon constitutive parameters calculated with the same homogenization model.  In Sec.~\ref{Sec_3} we apply this scattering model to calculate the scattering matrix of the interface between vacuum and a simple photonic crystal.  We then present two tests of the calculated scattering matrix elements.  Finally, in Sec.~\ref{Sec_4}, we use the results of the interface scattering model to postulate a set of phenomenological boundary conditions.

\section{Scattering matrix of the interface between a homogeneous medium and a metamaterial}\label{Sec_2}

\subsection{The constitutive parameters of a one dimensional metamaterial}

The one dimensional model that we present in this paper for calculating the scattering matrix of a metamaterial interface is strongly related to the one dimensional model used for metamaterial homogenization in Ref.~\onlinecite{Fietz_10b}.  The constitutive parameters used in this paper are calculated using the same homogenization method, originally presented for p-polarized waves in Ref.~\onlinecite{Fietz_10b}, and re-presented for s-polarized waves ($\textbf{E}=\mathrm{E}_z\hat{\textbf{z}}$ and $\textbf{H}=\mathrm{H}_x\hat{\textbf{x}}+\mathrm{H}_y\hat{\textbf{y}}$) in the appendix of this paper.  Here we briefly review the properties of the constitutive parameters determined from the one dimensional metamaterial homogenization model for s-polarized waves.

In this paper we will only consider two dimensional highly symmetric metamaterial/photonic crystals, though the concept is easily generalized to highly symmetric three dimensional crystals.  In two and three dimensions the symmetry requirements are that the crystal have a rectangular unit cell, and that the crystal has reflection symmetry in two directions tangential to the interface, making the direction normal to the interface a principle axis of the crystal.  In three dimensions, there is also the possibility of homogenizing a crystal with a rectangular unit cell as well as $180^{\circ}$ rotational symmetry around the direction normal to the interface, again making this direction a principle axis, even without reflection symmetry transverse to the interface.  In two and three dimensions, reflection symmetry in the direction of propagation (normal to the interface) is not necessary.  Also, for our limited two dimensional example, we are calculating scattering matrix elements for s-polarized waves, though it is straightforward to do the same for p-polarized waves.

The constitutive relationship for a s-polarized electromagnetic field that is harmonic in space and time, propagating in the $\hat{\textbf{x}}$-direction in a two dimensional crystal is 

\begin{equation}
\left(\!\!\begin{array}{c}
\mathrm{D}_z \\[5pt] \mathrm{B}_y
\end{array}\!\!\right) = 
\hat{C}\cdot
\left(\!\!\begin{array}{c}
\mathrm{E}_z \\[5pt] \mathrm{H}_y
\end{array}\!\!\right),
\end{equation}

\noindent where $\hat{C}$ is the constitutive matrix

\begin{equation}
\hat{C}(\omega,k_x) = 
\left(\!\!\!\begin{array}{cc}
\epsilon_{zz}(\omega,k_x) & \xi_{zy}(\omega,k_x) \\[5pt]
\zeta_{yz}(\omega,k_x) & \mu_{yy}(\omega,k_x)
\end{array}\!\!\!\right).
\end{equation}

\noindent  We emphasize that all of the constitutive parameters are functions of frequency $\omega$ and wavenumber $k_x$, the dependence on the wavenumber determining the degree of spatial dispersion, which in turn is the cause of the non-Maxwellian boundary conditions.

As shown in Ref.~\onlinecite{Fietz_10b}, if the metamaterial/photonic crystal of interest is reciprocal, the constitutive parameters obey the Lorentz reciprocity relations for a spatially dispersive material

\begin{equation}\label{Eq_reciprocity}
\begin{array}{ccc}
\epsilon_{zz}(\omega,-k_x) & = & \epsilon_{zz}(\omega,k_x), \\ \\ \mu_{yy}(\omega,-k_x) & = & \mu_{yy}(\omega,k_x), \\ \\ \xi_{zy}(\omega,-k_x) & = & -\zeta_{yz}(\omega,k_x).
\end{array}
\end{equation}

\noindent This allows us to represent the constitutive parameters as

\begin{equation}
\hat{C} = 
\left(\!\!\!\begin{array}{cc}
\epsilon_{zz} & \kappa_o+\kappa_e \\[5pt]
\kappa_o-\kappa_e & \mu_{yy}
\end{array}\!\!\!\right).
\end{equation}

\noindent Here $\epsilon_{yy}$, $\mu_{zz}$ and $\kappa_e$ are even functions of $k_x$ but $\kappa_o$ is an odd function of $k_x$.  In this paper we will only consider reciprocal metamaterial crystals.

From the constitutive parameters we can define an impedance for freely propagating waves

\begin{equation}\label{impedance}
\begin{array}{rl}
z^{\pm} = \displaystyle\mp\frac{\mathrm{E}_z}{\mathrm{H}_y} & = \displaystyle\frac{\omega\mu_{yy}/c}{k_x(\omega) + \omega\kappa_o/c\mp\omega\kappa_e/c} \\ \\
& = \displaystyle\frac{k_x(\omega)+\omega\kappa_o/c\pm\omega\kappa_e/c}{\omega\epsilon_{zz}/c}.
\end{array}
\end{equation}

\noindent Here $z^{\pm}$ is the impedance for a plane wave propagating in the positive (+) or negative (-) $\hat{\textbf{x}}$-direction, the difference between the two due to possible asymmetry of the crystal in the direction of propagation and the resulting nonzero $\kappa_e$.  The wavenumber $k_x(\omega)$ in Eq.~(\ref{impedance}) is the wavenumber of a freely propagating eigenmode of the crystal propagating in the positive $\hat{\textbf{x}}$-direction.  For a passive crystal this implies that $\mathrm{Im}(k_x)<0$ assuming the convention that harmonic plane waves vary as $e^{\mathrm{i}(\omega t-k_x x)}$.  The constitutive parameters in Eq.~(\ref{impedance}) are also evaluated using this positively propagating $k_x(\omega)$.

Finally, for a crystal with symmetry in the direction of propagation (here the $\hat{\textbf{x}}$ direction), symmetry forces $\kappa_e=0$.  Spatial dispersion however, still allows for nonzero bianisotropy through $\kappa_o$.

\subsection{Scattering model of the metamaterial interface}

\begin{figure}[t]
\begin{center}
\includegraphics[width=0.6\columnwidth]{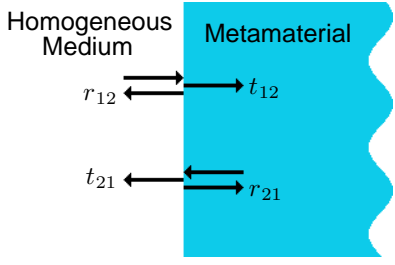}
\end{center}
\caption{A diagram of the scattering problem examined in this paper.  The scattering occurs at the interface between a semi-infinite homogeneous medium and a semi-infinte metamaterial.  The metamaterial is potentially nonlocal (spatially dispersive).}\label{Fig_1}
\end{figure}

The model presented in this section for calculating the scattering matrix of a metamaterial interface builds on a earlier model briefly described in Ref.~\onlinecite{Fietz_PHD} for calculating the reflection from a semi-infinite metamaterial.  This newer, more complete version of the model allows for the calculation of the entire scattering matrix of the metamaterial interface.  Fig.~\ref{Fig_1} provides a diagram of the metamaterial interface.  The incident waves are normally incident upon the interface, the normal direction being defined as the $\textbf{x}$-direction.  As before, the scattering matrix elements are labelled as $r_{ij}$ for electric field reflection amplitude of a plane wave incident upon the interface from medium $i$, and $t_{ij}$ for the electric field transmission amplitude of a plane wave incident upon the interface from medium $i$ and transmitted into medium $j$.  The homogeneous medium is labelled medium $1$ and the homogenized metamaterial is labelled medium $2$.

\begin{figure}[b!]
\begin{center}
\includegraphics[width=\columnwidth]{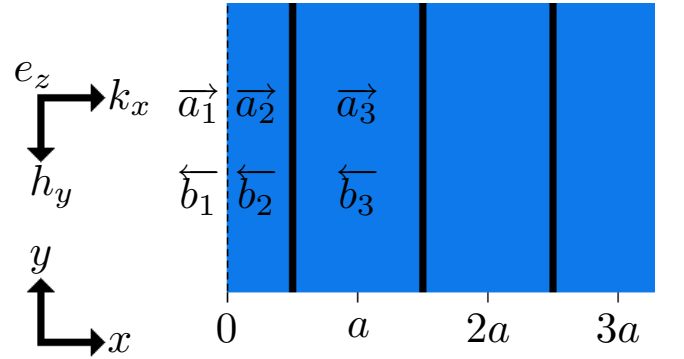}
\end{center}
\caption{A diagram of the simple one dimensional model used to solve the metamaterial interface scattering problem.  There are three domains, each with two s-polarized plane waves, one propagating in the positive $\hat{\textbf{x}}$ direction with electric field amplitude $a_i$, and one propagating in the negative $\hat{\textbf{x}}$ direction with electric field amplitude $b_i$.  The 1-st and 2-nd domains are separated by an interface between the homogeneous medium and the background (substrate) medium of the metamaterial.  The 2-nd and 3-rd domains are separated by a metasurface which serves as a subsitute for a single layer of the metamaterial crystal one unit cell thick.}\label{Fig_2}
\end{figure}

Just as in Ref.~\onlinecite{Fietz_10b}, we model the metamaterial crystal by replacing individual layers of the crystal with equivalent metasurfaces characterized by a surface polarizability.  The metasurfaces only interact with adjacent metasurfaces through planes waves.  Since all evanescent interactions are ignored, higher order propagating modes are absent from the model and the corresponding additional boundary conditions are not supported in the model.  As in Ref.~\onlinecite{Fietz_10b}, the one dimensional model of the interface has a algebraic solution which can easily be solved numerically.  There are six degrees of freedom in the model.  These are the coefficients $a_i$ and $b_i$, shown in Fig.~\ref{Fig_2}, which are the electric field amplitudes for plane waves travelling in the positive $\hat{\textbf{x}}$-direction ($a_i$) and negative $\hat{\textbf{x}}$-direction ($b_i$).  The microscopic electric and magnetic fields at any point in the domain of our one dimensional model are

\begin{equation}
\left(\begin{array}{cc}
\mathrm{e}_z \\[5pt] \mathrm{h}_y
\end{array}\right) = 
\left(\begin{array}{ccc}
e^{-\mathrm{i}k_ix} & \ & e^{\mathrm{i}k_i x} \\[5pt]
-\displaystyle\frac{1}{z_i}e^{-\mathrm{i}k_i x} & \ & \displaystyle\frac{1}{z_i}e^{\mathrm{i}k_i x}
\end{array}\right)\cdot
\left(\begin{array}{c}
a_i \\[5pt] b_i
\end{array}\right).
\end{equation}

\noindent Here $a_i$ and $b_i$ are the appropriate field coefficients for the particular value of $x$.  Also, $z_i=\sqrt{\mu_i/\epsilon_i}$ is the impedance and $k_i=\sqrt{\epsilon_i\mu_i}\omega/c$ is the wavenumber for the corresponding subdomain.  In general, the constitutive parameters of the homogeneous medium $\epsilon_1$ and $\mu_1$ can be different from the constitutive parameters of the background (substrate) of the metamaterial $\epsilon_i$ and $\mu_i$ for $i=2,3$.

  Only a single metasurface is necessary to represent the semi-infinite array.  It is trivial to include additional metasurfaces representing additional layers of the crystal, but this is unnecessary since it results in the same final interface scattering matrix.  Since there are six degrees of freedom, we require six equations of constraint.  These include four boundary conditions at the two interior boundaries (two for each boundary), as well as one boundary condition at each of the two exterior boundaries.  The first interior boundary to consider is the one between the homogeneous medium and the background medium (substrate) of the metamaterial located at $x=0$.  Due to Maxwellian boundary conditions at the interface between the vacuum and the background material (continuity of tangential electric and magnetic fields), the relationship between the 1-st and 2-nd sets of field coefficients is
  
\begin{equation}\label{Constraint_1}
\underbrace{\left(\begin{array}{ccc}
1 & \ & 1 \\[5pt]
-\displaystyle\frac{1}{z_1} & \ & \displaystyle\frac{1}{z_1}
\end{array}\right)}_{\hat{\mathrm{N}}_1}\cdot
\left(\begin{array}{c}
a_1 \\[5pt] b_1
\end{array}\right) = 
\underbrace{\left(\begin{array}{ccc}
1 & \ & 1 \\[5pt]
-\displaystyle\frac{1}{z_b} & \ & \displaystyle\frac{1}{z_b}
\end{array}\right)}_{\hat{\mathrm{N}}_2}\cdot
\left(\begin{array}{c}
a_2 \\[5pt] b_2
\end{array}\right).
\end{equation}

\noindent Next, we consider the relationship between the field coefficients of the second and third subdomains.  These two subdomains are separated by a metasurface.  This is an infinitesimally thin layer with the boundary condition

\begin{equation}\label{Delta_eq}
\left(\!\!\!\begin{array}{c}
\Delta\mathrm{h}_y \\[5pt] \Delta\mathrm{e}_z
\end{array}\!\!\!\right) = 
\mathrm{i}\displaystyle\frac{\omega}{c}
\left(\!\!\!\begin{array}{c}
\mathrm{p}_z \\[5pt] \mathrm{m}_y
\end{array}\!\!\!\right) = 
\mathrm{i}\displaystyle\frac{\omega}{c}\hat{\alpha}\cdot
\left(\!\!\!\begin{array}{c}
\mathrm{E}_z^{loc} \\[5pt] \mathrm{H}_y^{loc}
\end{array}\!\!\!\right).
\end{equation}

\noindent Here, $\Delta\mathrm{h}_y=\mathrm{h}_y^+-\mathrm{h}_y^-$, where $\mathrm{h}_y^+$ and $\mathrm{h}_y^-$ are the microscopic magnetic fields on the right and left sides of the metasurface respectively.  Similarly, $\Delta\mathrm{e}_z=\mathrm{e}_z^+-\mathrm{e}_z^-$ where $\mathrm{e}_z^+$ and $\mathrm{e}_z^-$ are the microscopic electric fields on the right and left sides of the metasurface respectively.  The local fields are defined as $\mathrm{E}_z^{loc}=(\mathrm{e}_z^++\mathrm{e}_z^-)/2$ and $\mathrm{H}_y^{loc}=(\mathrm{h}_y^++\mathrm{h}_y^-)/2$.  Finally, the electromagnetic response of the metasurface is characterized by the surface polarizability

\begin{equation}
\hat{\alpha}(\omega) = 
\left(\!\!\!\begin{array}{cc}
\alpha_{zz}^{ee} & \alpha_{zy}^{em} \\[5pt]
\alpha_{yz}^{me} & \alpha_{yy}^{mm}
\end{array}\!\!\!\right),
\end{equation}

\noindent which, can be derived from the numerically calculated scattering matrix of a single layer of the metasurface~\cite{Kuester_03,Fietz_10b}

\begin{widetext}
\begin{equation}\label{alpha_value}
\hat{\alpha}=-\displaystyle\frac{2\mathrm{i}}{\omega/c(1+S_{12}+S_{21}-\mathrm{det}(S))}\left(\!\!\!\begin{array}{cc}
\left[1+\mathrm{det}(S)-(S_{11}+S_{22})\right]/z_b & (S_{11}-S_{22})-(S_{12}-S_{21}) \\[5pt]
-(S_{11}-S_{22})-(S_{12}-S_{21}) & \left[1+\mathrm{det}(S)+(S_{11}+S_{22})\right]z_b
\end{array}\!\!\!\right).
\end{equation}

\noindent The scattering matrix of the metasurface in Eq.~(\ref{alpha_value}) is defined with the reference plane centered at the location of the metasurface.  We also note that Ref.~\onlinecite{Fietz_10b} presents a similar equation for the surface polarizability of a metasurface interacting with a p-polarized wave, as well as instructions for how to transform the p-polarized equation into an s-polarized version.  These, instructions were incorrect.  Eq.~(\ref{alpha_value}) is the correct equation for the surface polarizability of a metasurface interacting with a s-polarized wave.

Using Eq.~(\ref{Delta_eq}), we can relate the difference in the fields across the metasurface to the field coefficients with

\begin{equation}\label{M1}
\left(\!\!\!\begin{array}{c}
\Delta\mathrm{h}_y \\[5pt] \Delta\mathrm{e}_z
\end{array}\!\!\!\right) = 
\underbrace{\left(\!\!\!\begin{array}{cccc}
\displaystyle\frac{1}{z_b}e^{-\mathrm{i}k_ba/2} & -\displaystyle\frac{1}{z_b}e^{\mathrm{i}k_ba/2} & -\displaystyle\frac{1}{z_b}e^{-\mathrm{i}k_ba/2} & \displaystyle\frac{1}{z_b}e^{\mathrm{i}k_ba/2} \\[10pt]
-e^{-\mathrm{i}k_ba/2} & -e^{\mathrm{i}k_ba/2} & e^{-\mathrm{i}k_ba/2} & e^{\mathrm{i}k_ba/2}
\end{array}\!\!\!\right)}_{\hat{\mathrm{M}}_1}\cdot
\left(\!\!\begin{array}{c}
a_2 \\ b_2 \\ a_3 \\ b_3
\end{array}\!\!\right).
\end{equation}

\noindent Using the definition of the local fields, we can relate the local fields to the field coefficients with

\begin{equation}\label{M2}
\left(\!\!\begin{array}{c}
\mathrm{E}_z^{loc} \\[5pt] \mathrm{H}_y^{loc}
\end{array}\!\!\right) = 
\underbrace{\left(\!\!\!\begin{array}{cccc}
\displaystyle\frac{1}{2}e^{-\mathrm{i}k_ba/2} & \displaystyle\frac{1}{2}e^{\mathrm{i}k_ba/2} & \displaystyle\frac{1}{2}e^{-\mathrm{i}k_ba/2} & \displaystyle\frac{1}{2}e^{\mathrm{i}k_ba/2} \\[10pt]
-\displaystyle\frac{1}{2z_b}e^{-\mathrm{i}k_ba/2} & \displaystyle\frac{1}{2z_b}e^{\mathrm{i}k_ba/2} & -\displaystyle\frac{1}{2z_b}e^{-\mathrm{i}k_ba/2} & \displaystyle\frac{1}{2z_b}e^{\mathrm{i}k_ba/2}
\end{array}\!\!\!\right)}_{\hat{\mathrm{M}}_2}\cdot
\left(\!\!\begin{array}{c}
a_2 \\[3pt] b_2 \\[3pt] a_3 \\[3pt] b_3
\end{array}\!\!\right).
\end{equation}


\noindent Thus, the two boundary conditions in Eq.~(\ref{Delta_eq}) can be represented as 

\begin{equation}\label{Constraint_2}
\biggl(\hat{\mathrm{M}}_1-\mathrm{i}\displaystyle\frac{\omega}{c}\hat{\alpha}\cdot\hat{\mathrm{M}}_2\biggr)\cdot
\left(\!\!\begin{array}{c}
a_2 \\[3pt] b_2 \\[3pt] a_3 \\[3pt] b_3
\end{array}\!\!\right) = 
\left(\!\!\begin{array}{c}
0 \\ 0
\end{array}\!\!\right).
\end{equation}

\noindent If the fields associated with the field coefficients in Eq.~(\ref{Constraint_2}) are forced to be Bloch periodic, that is if the fields at $x=a$ differ from the fields at $x=0$ by the Bloch amplitude $e^{-\mathrm{i}k_xa}$, then the Bloch periodicity combined with the boundary conditions in Eq.~(\ref{Constraint_2}) allow us to derive a dispersion relation for the one dimensional array of metasurfaces

\begin{equation}\label{disp_rel}
\begin{array}{c}
\left(1+\displaystyle \frac{\omega^2}{c^2}\frac{\alpha_{zz}^{ee}\alpha_{yy}^{mm}-\alpha_{zy}^{em}\alpha_{yz}^{me}}{4}\right)\cos(k_xa)
 - \displaystyle\frac{\omega}{c}\frac{\alpha_{yz}^{me}+\alpha_{zy}^{em}}{2}\sin(k_xa)
 = \\ \\
\left(1-\displaystyle \frac{\omega^2}{c^2}\frac{\alpha_{zz}^{ee}\alpha_{yy}^{mm}-\alpha_{zy}^{em}\alpha_{yz}^{me}}{4}\right)\cos(k_0a) - \displaystyle\frac{\omega}{c}\frac{\alpha_{zz}^{ee}z_b+\alpha_{yy}^{mm}/z_b}{2}\sin(k_0a).
\end{array}
\end{equation}
\end{widetext}

\noindent This dispersion relation for s-polarized free waves propagating in a one dimensional array of metasurfaces is electromagnetically dual to the dispersion relation for p-polarized waves presented in Ref.~\onlinecite{Fietz_10b}.

We have now imposed four of the six required equations of constraint.  There remain two exterior boundary conditions that must be imposed in order to solve for the six field coefficients.  These two exterior boundary conditions involve the electric field amplitudes of plane waves that are incident upon the metamaterial interface.  First we consider a plane wave incident from vacuum.  The constraint for this is simply

\begin{equation}\label{Constraint_3}
a_1=\mathrm{E}_{h}^{inc},
\end{equation}

\noindent where $\mathrm{E}_h^{inc}$ is the electric field amplitude of the plane wave incident upon the interface from the homogeneous medium.  The constraint for a plane wave incident from the metamaterial is more complicated.  Eq.~(\ref{M2}) relates the field coefficients to the local fields, and the relation~\cite{Fietz_10b}

\begin{equation}\label{local_nonlocal_1}
\begin{array}{c}
\hat{\mathrm{Y}}\cdot
\left(\!\!\!\begin{array}{c}
\mathrm{E}_{m}^{out} \\[5pt] \mathrm{E}_{m}^{inc}
\end{array}\!\!\!\right) = 
\displaystyle\frac{\hat{\alpha}}{a}\cdot
\left(\!\!\begin{array}{c}
\textrm{E}_{z}^{loc} \\ \textrm{H}_{t}^{loc}
\end{array}\!\!\right),
\\ \\
\hat{\mathrm{Y}}=\Biggl(\hat{\chi}^+\cdot\left(\!\!\begin{array}{c} \scriptstyle 1 \\[1pt] \scriptstyle-\textstyle\frac{1}{z^+}\end{array}\!\!\right)e^{-\mathrm{i}k_xa/2}\raisebox{-10pt}{,} \ \ \ \ 
\hat{\chi}^-\cdot\left(\!\!\begin{array}{c} \scriptstyle 1 \\[1pt] \frac{1}{z^-}\end{array}\!\!\right)e^{\mathrm{i}k_xa/2}\Biggr),
\end{array}
\end{equation}


\noindent relates the local fields at the metasurface to $\mathrm{E}_m^{inc}$, the electric field amplitude of the plane wave incident upon the interface from the metamaterial and $\mathrm{E}_m^{out}$, the electric field amplitude of the plane wave propagating in the metamaterial outwards from the interface.  Both $\mathrm{E}_m^{inc}$ and $\mathrm{E}_m^{out}$ are electric field amplitudes at the location of the interface.  $\hat{\chi}^{\pm}$ is the macroscopic susceptibility defined in Eq.~(\ref{chi_express}) for plane waves moving in the positive ($\hat{\chi}^+=\hat{\chi}(\omega,k_x)$) and negative ($\hat{\chi}^-=\hat{\chi}(\omega,-k_x)$) $\hat{\textbf{x}}$-directions, where $k_x(\omega)$ is the wavenumber of a freely propagating eigenmode of the crystal moving in the positive $\hat{\textbf{x}}$-direction determined from Eq.~(\ref{disp_rel}).  Also, $z^{\pm}$ is the impedance of the homogenized metamaterial defined in Eq.~(\ref{impedance}) calculated with wavenumber $k_x(\omega)$ provided by the dispersion relation Eq.~(\ref{disp_rel}).  Both the left and right hand sides of Eq.~(\ref{local_nonlocal_1}) are equal to the polarization density of the homogenized metamaterial at the location of the metasurface.

Combining Eqs.~(\ref{M2}) and (\ref{local_nonlocal_1}), we get

\begin{equation}
\left(\!\!\begin{array}{c}
\mathrm{E}_m^{out} \\[5pt] \mathrm{E}_m^{inc}
\end{array}\!\!\right) = 
\underbrace{\hat{\mathrm{Y}}^{-1}
\cdot\displaystyle\frac{\hat{\alpha}}{a}\cdot\hat{M}_2}_{\hat{\mathrm{X}}}\cdot
\left(\!\!\begin{array}{c}
a_2 \\[3pt] b_2 \\[3pt] a_3 \\[3pt] b_3
\end{array}\!\!\right).
\end{equation}

\noindent This provides us with our final equation of constraint

\begin{equation}\label{Constraint_4}
(0,1)\cdot\hat{\mathrm{X}}\cdot\left(\!\!\begin{array}{c}
a_2 \\[3pt] b_2 \\[3pt] a_3 \\[3pt] b_3
\end{array}\!\!\right)
=\mathrm{E}_m^{inc}.
\end{equation}

We now have six equations of constraint to allow us to solve our model. Putting them all (Eqs.~(\ref{Constraint_1}),(\ref{Constraint_2}),(\ref{Constraint_3}) and (\ref{Constraint_4})) together, they are

\begin{equation}\label{Big_eq}
\left(\!\!\!\begin{array}{ccc}
\hat{\mathrm{N}}_1 & \ -\hat{\mathrm{N}}_2 & \begin{array}{cc} 0 & 0 \\ 0 & 0 \end{array} \\[8pt]
\begin{array}{cc} 0 & 0 \\ 0 & 0 \end{array} &  & \!\!\!\!\!\!\!\!\!\!\!\!\!\!\!\!\scriptstyle\hat{\mathrm{M}}_1-\mathrm{i}\textstyle\frac{\omega}{c}\scriptstyle\hat{\alpha}\cdot\hat{\mathrm{M}}_2 \\[8pt]
\begin{array}{cc} 1 & 0 \\ 0 & 0 \end{array} & \begin{array}{cc} 0 & 0 \\ & \end{array} & \begin{array}{cc} 0 & 0 \\ \!\!\!\!\!\!\!\!\!\!\!\!\!\!\! \scriptstyle(0,1)\cdot\hat{\mathrm{X}} & \end{array}
\end{array}\!\!\!\right)\cdot
\left(\!\!\begin{array}{c}
a_1 \\[3pt] b_1 \\[3pt] a_2 \\[3pt] b_2 \\[3pt] a_3 \\[3pt] b_3
\end{array}\!\!\right) = 
\left(\!\!\begin{array}{c}
0 \\[3pt] 0 \\[3pt] 0 \\[3pt] 0 \\[3pt] \mathrm{E}_h^{inc} \\[3pt] \mathrm{E}_m^{inc}
\end{array}\!\!\right).
\end{equation}

\noindent This small algebraic system of equations is easily solved numerically.

In order to calculate the scattering matrix for the metamaterial interface, we must solve Eq.~(\ref{Big_eq}) twice, once with a plane wave incident from the homogeneous medium, and once with a plane wave incident from the metamaterial.  First we solve the case of a plane wave incident from the homogeneous medium by setting $\mathrm{E}_m^{inc}=0$ and setting $\mathrm{E}_h^{inc}$ to an arbitrary non-zero value.  After solving Eq.~(\ref{Big_eq}) for the field coefficients, we can calculate two of the scattering matrix elements of the interface,

\begin{equation}
\begin{array}{rl}
r_{12} & = \displaystyle\frac{b_1}{\mathrm{E}_v^{inc}}, \\
t_{12} & = \displaystyle\frac{1}{\mathrm{E}_v^{inc}}
(1,0)\cdot\hat{\mathrm{X}}\cdot
\left(\!\!\begin{array}{c}
a_2 \\[3pt] b_2 \\[3pt] a_3 \\[3pt] b_3
\end{array}\!\!\right).
\end{array}
\end{equation}

\noindent Next, we solve the case of a plane wave incident from the metamaterial by setting $\mathrm{E}_h^{inc}=0$ and setting $\mathrm{E}_m^{inc}$ to an arbitrary non-zero value.  After solving Eq.~(\ref{Big_eq}) for the field coefficients, the remaining two scattering matrix elements are

\begin{equation}
\begin{array}{rl}
r_{21} & = \displaystyle\frac{1}{\mathrm{E}_m^{inc}}
(1,0)\cdot\hat{\mathrm{X}}\cdot
\left(\!\!\begin{array}{c}
a_2 \\[3pt] b_2 \\[3pt] a_3 \\[3pt] b_3
\end{array}\!\!\right), \\
t_{21} & = \displaystyle\frac{b_1}{\mathrm{E}_m^{inc}}.
\end{array}
\end{equation}

\section{Example: Scattering matrix of a photonic crystal interface}\label{Sec_3}

\subsection{Dispersion relation and scattering matrix}

\begin{figure}[t]
\begin{center}
\includegraphics[width=\columnwidth]{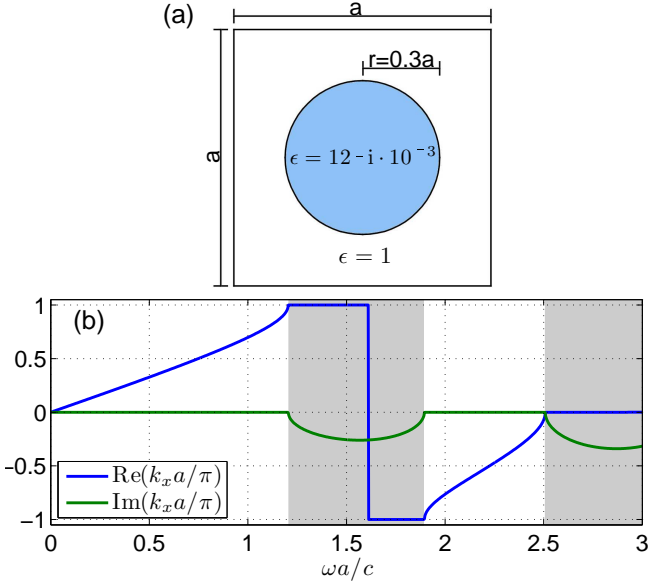}
\end{center}
\caption{(a) Diagram of the unit cell of the photonic crystal examined in this section.  (b) Complex $k_x$ dispersion relation for an s-polarized eigenmode propagating along a principle axis of the photonic crystal.  Band gaps are indicated by nonzero $\mathrm{Im}(k_x)$ and are shaded.}\label{Fig_3}
\end{figure}

As a demonstration of the metamaterial interface scattering model, we calculate the scattering matrix of the interface between vacuum and a simple two dimensional photonic crystal, the unit cell of which is shown in Fig.~\ref{Fig_3}.  The unit cell is square with lattice constant $a$.  At the center of the unit cell is a cylinder with radius $r=0.3a$.  The cylinder is a dielectric with permittivity $\epsilon=12-\mathrm{i}\cdot 10^{-3}$ and is surrounded by vacuum with permittivity $\epsilon=1$.  The negative imaginary part in the permittivity of the cylinder indicates loss.  Fig.~\ref{Fig_3} also shows a complex $k_x(\omega)$ dispersion relation for a one dimensional array of metasurfaces corresponding to the cylindrical photonic crystal and calculated with Eq.~(\ref{disp_rel}).  Notice there are two band gaps, indicated by the non-zero imaginary part of $k_x$ and shaded in Fig.~\ref{Fig_3} and in the remaining figures.  This dispersion relation agrees very well with the complex $k_x(\omega)$ dispersion relation of the cylindrical photonic crystal calculated from a finite element simulation~\cite{Marcelo_07} (not shown), except for a narrow band near $\omega=2.5c/a$.

The constitutive parameters of the photonic crystal must first be determined if we are to calculate the scattering matrix of the interface.  This is done with the procedure presented for p-polarized waves in Ref.~\onlinecite{Fietz_10b}, and re-presented in the appendix for s-polarized waves.  First, the surface polarizability of a single layer of the crystal is calculated from the scattering matrix of a single layer.  This is then used to calculate a set of macroscopic constitutive parameters relating the electric and magnetic fields to the electric displacement and magnetic flux density fields

\begin{equation}
\left(\!\!\begin{array}{c}
\mathrm{D}_z \\[5pt] \mathrm{B}_y
\end{array}\!\!\right) = 
\left(\!\!\begin{array}{cc}
\epsilon_{zz} & \kappa_o \\[5pt] \kappa_o & \mu_{yy}
\end{array}\!\!\right)\cdot
\left(\!\!\begin{array}{c}
\mathrm{E}_z \\[5pt] \mathrm{H}_y
\end{array}\!\!\right).
\end{equation}

\begin{figure}[t]
\begin{center}
\includegraphics[width=\columnwidth]{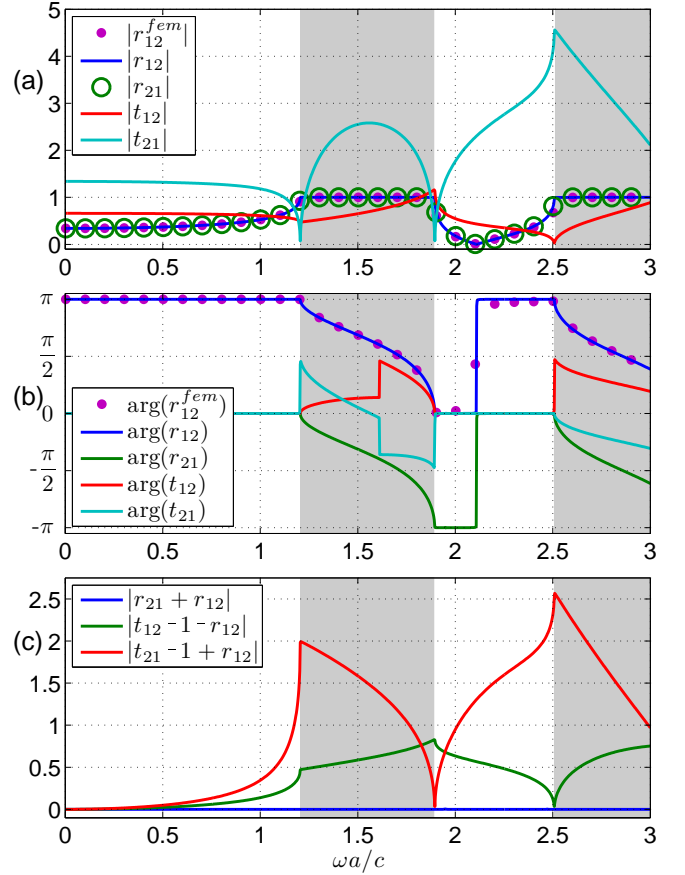}
\end{center}
\caption{(a) Absolute value and (b) argument of the reflection amplitude $r_{12}^{fem}$ calculated from a finite element simulation as well as the scattering matrix elements $r_{12}$, $r_{21}$, $t_{12}$ and $t_{21}$ of the vacuum/crystal interface calculated from the model presented in Sec.~\ref{Sec_2}B.  (c) Test of the relations in Eq.~(\ref{refl_2}).  Shaded regions indicate a band gap.  The relation $r_{12}=-r_{21}$ appears to be valid while the relations in Eq.~(\ref{refl_2}) for the transmission coefficients fail.  Notice that the scattering matrix elements are real valued outside the band gap but complex valued inside the band gap.  Also, in the band gap $\vert r_{12} \vert = \vert r_{21} \vert = 1$.}\label{Fig_4}
\end{figure}

\begin{figure*}[t]
\begin{center}
\includegraphics[width=\textwidth]{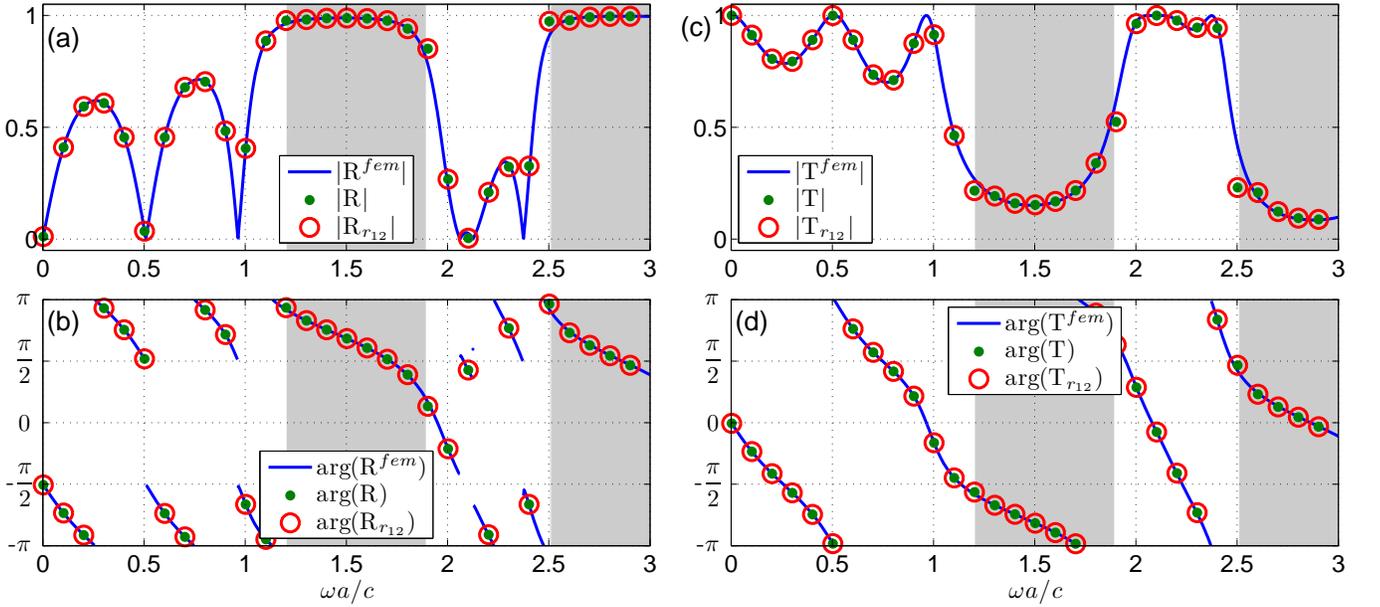}
\end{center}
\caption{Comparision of reflection and transmission amplitudes for a photonic crystal slab three units cell thick, calculated in three ways.  From a finite element simulation ($\mathrm{R}^{fem}$ and $\mathrm{T}^{fem}$), from Eq.~(\ref{RT}) using all four scattering matrix elements calculated from the model ($\mathrm{R}$ and $\mathrm{T}$), and from the simplified expressions in Eq.~(\ref{RT_r12}) calculated from $r_{12}$ ($\mathrm{R}_{r_{12}}$ and $\mathrm{T}_{r_{12}}$).  (a) Absolute value and (b) argument of the reflection amplitudes.  (c) Absolute value and (d) argument of the transmission amplitudes.  Shaded regions indicate a band gap.  Note that the apparent $\pi$ jumps in $arg(\mathrm{R})$ occur when $\vert \mathrm{R} \vert$ approaches zero.}\label{Fig_5}
\end{figure*}

\noindent  Each of these constitutive parameters is a function of both $\omega$ and $k_x$, the dependence on $k_x$ indicating spatial dispersion.  Since we are only considering s-polarized waves, and due to the symmetry and reciprocity of the crystal unit cell, we only need to consider three constitutive parameters.  Also, as mentioned in Sec.~\ref{Sec_2}A, because of the reciprocity of the crystal, both $\epsilon_{zz}$ and $\mu_{yy}$ are even functions of $k_x$, and $\kappa_o$ is an odd function of $k_x$.  Using these constitutive parameters, along with the model described in the previous section, we can calculate the scattering matrix for the interface between vacuum and the photonic crystal.  The four parameters of the scattering matrix are plotted in Fig.~\ref{Fig_4}.

The first important result in Fig.~\ref{Fig_4} is the agreement between $r_{12}^{fem}$, which is the the reflection amplitude of a semi-infinite photonic crystal calculated from a finite element simulation and $r_{12}$, which is a scattering matrix element calculated from the model presented in Sec.~\ref{Sec_2}B.  It is not known how to calculate the other scattering matrix elements from a finite element simulation and therefore it is not possible to test them in this way.  The second result, is that the two reflection coefficients $r_{12}$ and $r_{21}$ do in fact appear to be the negative of each other.  This can be seen more clearly in Fig.~\ref{Fig_4}c, which plots the values $\vert r_{21}+r_{12} \vert$, $\vert t_{12}-1-r_{12} \vert$ and $\vert t_{21}-1+r_{12} \vert$.  If the relations in Eq.~(\ref{refl_2}) are correct then the expressions plotted in Fig.~\ref{Fig_4}c should all be zero.  We see from Fig.~\ref{Fig_4}c that this indeed true for the relation $r_{21}=-r_{12}$.  However, the two remaining relations in Eq.~(\ref{refl_2}), while valid in the long wavelength limit, do appear to fail as expected due to spatial dispersion.

\subsection{Test: Transmission and reflection of a photonic crystal slab}

As a test of the calculated interface scattering matrix, we can use the scattering matrix elements to calculate the reflection and transmission through a photonic crystal slab.  The calculated reflection and transmission amplitudes can then be compared against values calculated from a finite element simulation of the photonic crystal slab.  For a photonic crystal slab, three layers thick and with two boundaries with vacuum, each boundary described by scattering matrix elements calculated from our model, the total reflection and transmission amplitudes for the slab are

\begin{equation}\label{RT}
\begin{array}{rl}
\mathrm{R} = & r_{12} + \displaystyle\frac{t_{12}r_{21}t_{21}e^{-6\mathrm{i}k_xa}}{1-r_{21}^2e^{-6\mathrm{i}k_xa}}, \\ \\
\mathrm{T} = & \displaystyle\frac{t_{12}t_{21} e^{-3\mathrm{i}k_xa}}{1-r_{21}^2e^{-6\mathrm{i}k_xa}}.
\end{array}
\end{equation}

\noindent Here $k_x(\omega)$ is a complex valued wavenumber calculated from Eq.~(\ref{disp_rel}) and plotted in Fig.~\ref{Fig_3}.  Notice that we have not used any of the relations in Eq.~(\ref{refl_2}) which are often used to simplify these types of expressions.  If we had used Eq.~(\ref{refl_2}), the reflection and transmission amplitudes could have been expressed as

\begin{equation}\label{RT_r12}
\begin{array}{rl}
\mathrm{R}_{r_{12}} = & r_{12} \displaystyle\frac{1-e^{-6\mathrm{i}k_xa}}{1-r_{12}^2e^{-6\mathrm{i}k_xa}}, \\ \\
\mathrm{T}_{r_{12}} = & \displaystyle\frac{1-r_{12}^2}{1-r_{12}^2e^{-6\mathrm{i}k_xa}}e^{-3\mathrm{i}k_xa}.
\end{array}
\end{equation}

In Fig.~\ref{Fig_5} we have plotted the complex valued reflection amplitudes $\mathrm{R}^{fem}$, $\mathrm{R}$ and $\mathrm{R}_{r_{12}}$, and the transmission amplitudes $\mathrm{T}^{fem}$, $\mathrm{T}$ and $\mathrm{T}_{r_{12}}$.  Here $\mathrm{R}^{fem}$ and $\mathrm{T}^{fem}$ are calculated from a finite element simulation of a three layered photonic crystal slab.

We see from Fig.~\ref{Fig_5} that the transmission and reflection amplitudes calculated using Eq.~(\ref{RT}) with the scattering matrix elements of the vacuum-crystal interface calculated from our model agree very well with the finite element simulation.  However, the simplified reflection and transmission amplitudes of Eq.~(\ref{RT_r12}) using only $r_{12}$ also agree with the finite element simulation.  This is a surprise since we have seen in Fig.~\ref{Fig_4} that with the exception of $r_{21}=-r_{12}$, the remaining relations of Eq.~(\ref{refl_2}) fail due to spatial dispersion.  Comparing Eqs.~(\ref{RT}-\ref{RT_r12}), we find a new relation between the scattering matrix elements,

\begin{equation}\label{sym_rel_2}
t_{12}t_{21}=(1+r_{12})(1+r_{21}),
\end{equation}

\noindent  which is easily confirmed by directly examining the scattering matrix elements returned by our model.  This new relation is true despite the fact that the relations for the transmission amplitudes in Eq.~(\ref{refl_2}) fail due to spatial dispersion.

\subsection{Test: Poynting flux at the vacuum-crystal interface}

There is an additional test we can perform to test the accuracy of the calculated scattering matrix elements of the vacuum-crystal interface.  Energy flux at the interface must be conserved.  If we use the calculated scattering matrix elements to calculate the macroscopic energy flux and compare it to the true microscopic energy flux, this provides us with a strong test of the scattering matrix elements.  At the same time, calculating the energy flux is a good test of the nonlocal homogenization method we have used~\cite{Fietz_10b} to calculate the constitutive parameters of the photonic crystal.  This homogenization procedure deliberately calculates constitutive parameters that are functions of the wavenumber.  This is essential for correctly calculating the Poynting flux since the term $\textbf{S}^1$ in Eq.~(\ref{Poynt_3}) involves partial derivatives of the constitutive parameters with respect to the wavenumber.  To the best knowledge of the authors, there has been only one other peer reviewed paper where the nonlocal Poynting flux has been calculated inside of a metamaterial~\cite{Costa_11}.

We can calculate the Poynting flux at the vacuum-crystal interface for two different scenarios. First, where a plane wave is incident upon the interface from the vacuum and second, when a plane wave is incident upon the interface from the crystal.  In the first case where the plane wave is incident from the vacuum, the macroscopic Poynting flux in the $\hat{\textbf{x}}$ direction is


\begin{equation}\label{PS1}
\begin{array}{l}
\mathrm{S}_v =  \mathrm{S}_v^0+\mathrm{S}_v^1, \\ \\
\mathrm{S}_v^0 =  -\displaystyle\frac{\mathrm{Re}[\mathrm{E}_z\mathrm{H}_y^*]}{2} = \frac{1}{2}\mathrm{Re}\left[\frac{1}{z^+}\right]\vert t_{12} \vert^2 \vert \mathrm{E}_v^{inc} \vert^2, \\ \\
\mathrm{S}_v^1 =  \displaystyle\frac{1}{2}\mathrm{Re}\left[\Xi^+\right]\vert t_{12} \vert^2 \vert \mathrm{E}_v^{inc} \vert^2,
\end{array}
\end{equation}

\noindent where

\begin{equation}\label{PS2}
\begin{array}{rl}
\Xi^{\pm} = \displaystyle-\frac{\omega}{2}\biggl( \biggr. \!\!\!& \left.\displaystyle \frac{\partial\epsilon}{\partial k_x} - 2\mathrm{Re}\left[\frac{1}{z^{\pm}}\right]\frac{\partial\kappa_o}{\partial k_x} \right. \\ \\
& \biggl.\displaystyle \mp 2\mathrm{i}\mathrm{Im}\left[\frac{1}{z^{\pm}}\right]\frac{\partial\kappa_e}{\partial k_x} + \frac{1}{\vert z^{\pm}\vert^2}\frac{\partial\mu}{\partial k_x} \biggr).
\end{array}
\end{equation}

\noindent Note that the derivatives of the constitutive parameters are evaluated for real valued wavenumbers $k_x$.

In the second scenario, with a plane wave incident from the metamaterial crystal, there are now two waves in the crystal, one incident and one reflected.  For the case of two overlapping plane waves, each wave has a corresponding $\textbf{S}_1$ (Eq.~(\ref{Poynt_3})) correction to the Poynting flux.  This requires that the derivatives of the constitutive parameters be evaluated for both positive $k_x$, for the reflected wave moving in the positive $\hat{\textbf{x}}$-direction, as well as evaluated for negative $k_x$, for the incident wave moving in the negative $\hat{\textbf{x}}$-direction.  Assuming Lorentz reciprocity, ensuring that $\epsilon_{zz}$, $\mu_{yy}$ and $\kappa_e$ are even in $k_x$ and $\kappa_o$ is odd in $k_x$, we can relate the derivatives of the constitutive parameters evaluated at negative $k_x$, to the derivatives evaluated at positive $k_x$.  Done correctly, the Poynting flux in the $\hat{\textbf{x}}$ direction at the vacuum-crystal interface with a plane wave incident from the crystal is


\begin{equation}\label{PS3}
\begin{array}{rl}
\mathrm{S}_m = & \mathrm{S}_m^0+\mathrm{S}_m^1, \\ \\
\mathrm{S}_m^0 = & -\displaystyle\frac{\mathrm{Re}[\mathrm{E}_z\mathrm{H}_y^*]}{2} \\ \\
= & \displaystyle\frac{1}{2}\mathrm{Re}\left[ \displaystyle \frac{\vert r_{21} \vert^2}{z^+} + \frac{r_{21}}{z^+} - \frac{r_{21}^*}{z^-} - \frac{1}{z^-} \right]\vert \mathrm{E}_m^{inc} \vert^2, \\ \\
\mathrm{S}_m^1 = & \displaystyle\frac{1}{2}\biggl(\mathrm{Re}\left[\Xi^+\right]\vert r_{21} \vert^2 - \mathrm{Re}\left[\Xi^-\right] \biggr) \vert \mathrm{E}_m^{inc} \vert^2.
\end{array}
\end{equation}

\noindent It should be noted that the derivation for Eq.~(\ref{Poynt_3})~\cite{Agranovich_Energy,Landau_Energy,Kamenetskii_96} assumes that the frequency $\omega$ and wavenumber $k_x$ are approximately real valued.  This is not necessarily true for an evanescent wave or for a propagating wave in the presence of high losses.  In either case the expression for the Poynting flux in Eqs.~(\ref{PS1}-\ref{PS3}) is in general not correct.

\begin{figure}[t]
\begin{center}
\includegraphics[width=\columnwidth]{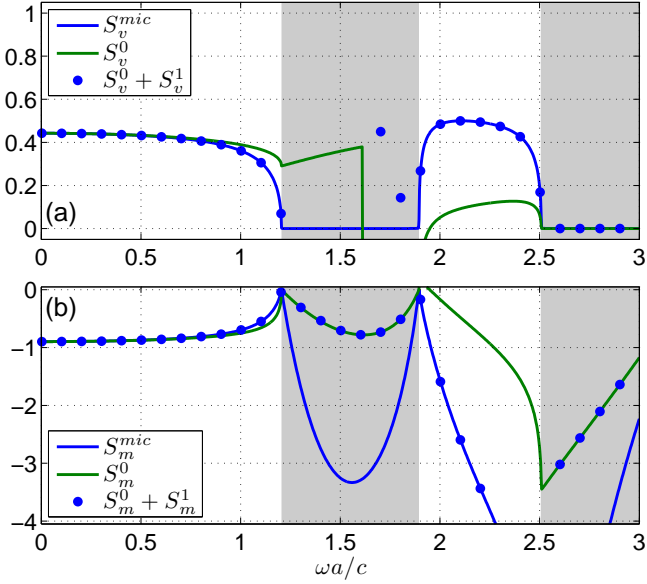}
\end{center}
\caption{Poynting fluxes in a spatialy dispersive medium (homogenized photonic crystal) at the metamaterial interface for (a) plane wave incident from the vacuum and (b) plane wave incident from the photonic crystal.  Shaded regions indicate a band gap.  In both cases the incident electric field is normalized to unity.  We see good agreement between the spatially dispersive Poynting flux expressions $\mathrm{S}=\mathrm{S}^0+\mathrm{S}^1$ and the microscopic Poynting flux $\mathrm{S}^{mic}$ for frequencies outside the band gap.}\label{Fig_6}
\end{figure}

The two Poynting fluxes $\mathrm{S}_v$ and $\mathrm{S}_m$ are plotted in Fig~\ref{Fig_6}, along with the microscopic Poynting flux in the $\hat{\textbf{x}}$ direction for each case

\begin{equation}
\mathrm{S}^{mic} = -\displaystyle\frac{1}{2}\mathrm{Re}[\mathrm{e}_z\mathrm{h}_y^*].
\end{equation}

\noindent By definition, the microscopic fields at the interface agree with the macroscopic fields on the vacuum side of the interface.  In both cases, the incident electric field is normalized to unity.  We can see from Fig.~\ref{Fig_6}, that the Poynting flux calculated from the scattering matrix elements of the metamaterial interface and the constitutive parameters of the homogenized photonic crystal agrees very well with the microscopically calculated Poynting flux for frequencies that lie in the passband.  For frequencies in the bandgap, there is very little agreement between the macroscopic and microscopic Poynting fluxes.  As mentioned earlier, this is due to the large imaginary part of wavenumber $k_x$ of an evanescent wave in a band gap.  However, the fact that we get the correct Poynting flux in the passband is positive evidence that the calculated scattering matrix elements are correct in the pass band.

Both tests of the calculated interface scattering matrix elements returned positive results, however it should be noted that the first test in Sec.~\ref{Sec_3}B really indicates the validity of the relationships $r_{21}=-r_{12}$ and $t_{12}t_{21}=(1+r_{12})(1+r_{21})$ between the various scattering matrix elements.  Similarly, the test presented in Sec.~\ref{Sec_3}C only tests the absolute value of the scattering matrix elements, and only provides positive results in the passbands since Eqs.~(\ref{PS1}-\ref{PS3}) are only valid in the passbands.

\section{Phenomenological non-Maxwellian boundary conditions}\label{Sec_4}

\subsection{Interface between local homogeneous medium and a symmetric crystal}

A surprising result of our investigation of the scattering matrix of the photonic crystal interface in Fig.~\ref{Fig_3} was that the relation $r_{12}=-r_{21}$ remains true even when spatial dispersion forces the remaining relations of Eq.~(\ref{refl_2}) to fail.  Our test of the total reflection and transmission amplitudes for a slab of the cylindrical metamaterial led to the discovery of another relation in Eq.~(\ref{sym_rel_2}).

\begin{figure}[t]
\begin{center}
\includegraphics[width=\columnwidth]{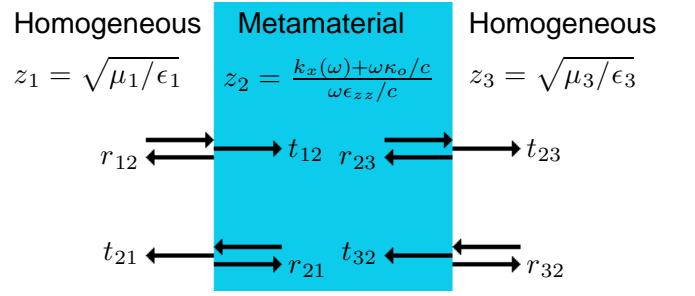}
\end{center}
\caption{Diagram of a slab of a nonlocal homogenized metamaterial surrounded on both sides with semi-infinite local homogeneous media with differing constitutive parameters.}\label{Fig_7}
\end{figure}

In an attempt to try to generalize these relations, and use them to infer some insights into non-Maxwellian boundary conditions for metamaterials with spatial dispersion, we consider measuring the total reflection and transmission through a metamaterial slab, only this time the homogeneous media on the opposite sides of the slab have different constitutive parameters than vacuum and from each other.  This simple geometry is shown in Fig.~\ref{Fig_7}.

Assuming that the three different media are lossless and that the metamaterial has reflection symmetry in the $\hat{\textbf{x}}$-direction ($\kappa_e=0$), and by performing the simple analytic derivation of the total reflection and transmission amplitudes of the slab, it is easy to see that conservation of energy can be guaranteed if the following relations are true

\begin{equation}\label{sym_rel_3}
\begin{array}{c}
\displaystyle\frac{1}{z_1}\left(\frac{1-r_{12}}{1+r_{12}}\right)=\frac{1}{z_3}\left(\frac{1-r_{32}}{1+r_{32}}\right), \\ \\
t_{12}t_{23}=\left(1+r_{12}\right)\left(1+r_{23}\right).
\end{array}
\end{equation}

\noindent These relations can be easily confirmed numerically with the model presented in Sec.~\ref{Sec_2}B.

The relation $r_{12}=-r_{21}$, along with Eqs.~(\ref{sym_rel_2},\ref{sym_rel_3}) allow us to guess the form of the boundary conditions for the interface between a local homogeneous medium and a nonlocal homogenized metamaterial.  For the interface in Fig.~\ref{Fig_7} between media 1 and 2, the boundary conditions are

\begin{equation}\label{bound_con_1}
\begin{array}{rl}
\underbrace{1+r_{12}}_{\mathrm{E}_z^1} =& \underbrace{at_{12}}_{a\mathrm{E}_z^2}, \\ \\
\displaystyle \underbrace{-\frac{1}{z_1}+\frac{r_{12}}{z_1}}_{\mathrm{H}_y^1} =& \underbrace{-b\frac{t_{12}}{z_2}}_{b\mathrm{H}_y^2}, \\ \\
\underbrace{t_{21}}_{\mathrm{E}_z^{1}} =& \underbrace{a\left(r_{21}+1\right)}_{a\mathrm{E}_z^{2}}, \\ \\
\underbrace{\displaystyle\frac{t_{21}}{z_1}}_{\mathrm{H}_y^{1}} =& \underbrace{b\left(-\frac{r_{21}}{z_2}+\frac{1}{z_2}\right)}_{b\mathrm{H}_y^{2}}.
\end{array}
\end{equation}

\begin{figure}[h]
\begin{center}
\includegraphics[width=\columnwidth]{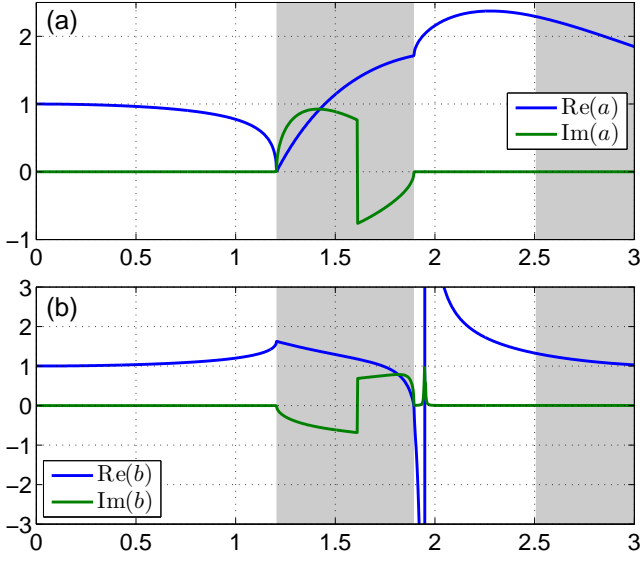}
\end{center}
\caption{The interface parameters of Eq.~(\ref{bound_con_1}) for the nonlocal medium corresponding to the cylindrical photonic crystal shown in Fig.~\ref{Fig_3}.}\label{Fig_8}
\end{figure}

\noindent  Here $\mathrm{E}_z^1$, $\mathrm{E}_z^2$, $\mathrm{H}_y^1$ and $\mathrm{H}_y^2$ are the electric and magnetic field amplitudes in media 1 and 2, and in every case the incident electric field is normalized to 1.  The coefficient a and b are interface parameters characterizing the non-Maxwellian boundary condition at the interface.  These phenomenological boundary conditions are also easily confirmed numerically with the scattering model presented in Sec.~\ref{Sec_2}B, both for lossless and lossy metamaterials.  The interface parameters $a$ and $b$ for the cylindrical photonic crystal shown in Fig.~\ref{Fig_3} are plotted in Fig.~\ref{Fig_8}.

There are a number of conclusions we can draw from Eq.~(\ref{bound_con_1}).  First, as can be seen in Fig.~\ref{Fig_8}, in the long wavelength limit, both $a$ and $b$ equal 1, corresponding to the standard Maxwellian boundary conditions.  Second, the discontinuity in the transverse electric ($\mathrm{E}_z$) and magnetic ($\mathrm{H}_y$) fields is always proportional to the electromagnetic field strength on the metamaterial side of the interface.  The field strength on the side of the interface with the local homogeneous medium has no effect on the field discontinuities.  Furthermore, the discontinuity in the magnetic field, which is usually associated with an electric surface current, is proportional to the magnetic field.  Likewise, the discontinuity in the electric field, normally associated with a magnetic surface current, is proportional to the electric field.  If one tried to characterize the metamaterial interface as a metasurface, as was attempted in Ref.~\onlinecite{Kim_11}, according to Eq.~(\ref{bound_con_1}), the surface currents would be proportional to the electromagnetic fields on the metamaterial side of the interface, and the surface polarizability matrix would have zeros along its diagonal and nonzero values for its off diagonal components.  Lastly, though this is not evident from Eq.~(\ref{bound_con_1}), it can be demonstrated numerically using the model presented in Sec.~\ref{Sec_2}B that the interface parameters $a$ and $b$ only depend on the particular metamaterial.  They are unchanged when the medium bordering the metamaterial is changed.  We will return to this point in Sec.~\ref{Sec_4}B.

\subsection{Interface between local homogeneous medium and an asymmetric crystal}

So far we have only considered crystals that have reflection symmetry in the direction of wave propagation.  We now briefly consider a crystal that does not have reflection symmetry in the propagation direction, though it does still have reflection symmetries in the directions perpendicular to propagation.  We cannot expect the relations we postulated in Eq.~(\ref{bound_con_1}) for a symmetric crystal to be correct for an asymmetric crystal.  It is easy to see that in the long wavelength limit, we cannot even expect the relation $r_{12}=-r_{21}$ to be valid for an asymmetric crystal, due to the fact that in an asymmetric crystal the impedance depends on weather the electromagnetic wave is propagating in the positive or negative $\hat{\textbf{x}}$-direction (see Eq.~(\ref{impedance})).

There are a few thing we can assume about the boundary conditions for an asymmetric crystal.  First, in the limit of an asymmetric crystal becoming symmetric, $\kappa_e$ vanishes, $z^+=z^-$, and the boundary conditions should reduce to Eq.~(\ref{bound_con_1}).  Second, by examining Eqs.~(\ref{PS1}-\ref{PS3}), we can see that any coefficients determining the boundary conditions for the interface with an asymmetric crystal, analogous to $a$ and $b$ for a symmetric crystal, should depend on the direction of propagation of the electromagnetic waves.  With these insights, we can guess the form of the non-Maxwellian boundary conditions for the interface of an asymmetric crystal.  For an asymmetric crystal, that when homogenized becomes nonlocal and is labelled as medium 2, surrounded by two local homogeneous media labelled 1 and 3, the boundary conditions for the interface between media 1 and 2 are

\begin{equation}\label{bound_con_2}
\begin{array}{rl}
\underbrace{1}_{\mathrm{E}_z^{1+}}+\underbrace{r_{12}}_{\mathrm{E}_z^{1-}} =& \underbrace{a^+t_{12}}_{a^+\mathrm{E}_z^{2+}}, \\ \\
\displaystyle \underbrace{-\frac{1}{z_1}}_{\mathrm{H}_y^{1+}}+\underbrace{\frac{r_{12}}{z_1}}_{\mathrm{H}_y^{1-}} =& \underbrace{-b^+\frac{t_{12}}{z_2^+}}_{b^+\mathrm{H}_y^{2+}}, \\ \\
\underbrace{t_{21}}_{\mathrm{E}_z^{1-}} =& \underbrace{a^+r_{21}}_{a^+\mathrm{E}_z^{2+}}+\underbrace{a^-}_{a^-\mathrm{E}_z^{2-}}, \\ \\
\underbrace{\displaystyle\frac{t_{21}}{z_1}}_{\mathrm{H}_y^{1-}} =& \underbrace{-b^+\frac{r_{21}}{z_2^+}}_{b^+\mathrm{H}_y^{2+}} + \underbrace{b^-\frac{1}{z_2^-}}_{b^-\mathrm{H}_y^{2-}},
\end{array}
\end{equation}

\noindent and  the boundary conditions for the interface between media 2 and 3 are

\begin{equation}\label{bound_con_3}
\begin{array}{rl}
\underbrace{a^+}_{a^+\mathrm{E}_z^{2+}}+\underbrace{a^-r_{23}}_{a^-\mathrm{E}_z^{2-}} =& \underbrace{t_{23}}_{\mathrm{E}_z^{3+}}, \\ \\
\displaystyle \underbrace{-b^+\frac{1}{z_2^+}}_{b^+\mathrm{H}_y^{2+}}+\underbrace{b^-\frac{r_{23}}{z_2^-}}_{b^-\mathrm{H}_y^{2-}} =& \underbrace{-\frac{t_{23}}{z_3}}_{\mathrm{H}_y^{3+}}, \\ \\
\underbrace{a^-t_{32}}_{a^-\mathrm{E}_z^{2-}} =& \underbrace{r_{32}}_{\mathrm{E}_z^{3+}}+\underbrace{1}_{\mathrm{E}_z^{3-}}, \\ \\
\underbrace{\displaystyle b^-\frac{t_{32}}{z_2^-}}_{b^-\mathrm{H}_y^{2-}} =& \underbrace{-\frac{r_{32}}{z_3}}_{\mathrm{H}_y^{3+}} + \underbrace{\frac{1}{z_3}}_{\mathrm{H}_y^{3-}}.
\end{array}
\end{equation}

\noindent Here $z_1$ and $z_3$ are the impedances of media 1 and 3, and $z_2^{\pm}$ are the impedances of the asymmetric homogenized metamaterial given by Eq.~(\ref{impedance}).  $\mathrm{E}_z^{1\pm}$, $\mathrm{E}_z^{2\pm}$ and $\mathrm{E}_z^{3\pm}$ and $\mathrm{H}_y^{1\pm}$, $\mathrm{H}_y^{2\pm}$ and $\mathrm{H}_y^{3\pm}$ are the electric and magnetic field amplitudes in media 1,2 and 3 for waves propagating in a positive ($+$) or negative ($-$) $\hat{\textbf{x}}$-directions, and in every case the incident electric field is normalized to 1.  Finally, $a^{\pm}$ and $b^{\pm}$ are interface parameters for waves propagating in the positive ($+$) and negative ($-$) $\hat{\textbf{x}}$-directions.

As with the interface parameters for the symmetric metamaterial, the parameters $a^{\pm}$ and $b^{\pm}$ depend solely on the particular metamaterial and are not affected by the medium on the opposite side of the interface.  Similarly, the discontinuity in the tangential electric and magnetic fields is entirely due to the electromagnetic field strength on the metamaterial side of the interface.  This can all easily be confirmed from the interface scattering model presented in Sec.~\ref{Sec_2}B.

Finally, though the interface scattering model presented in Sec.~\ref{Sec_2}B is for an interface between a local homogeneous medium and a nonlocal homogenized metamaterial, this model can easily be modified to calculate the scattering matrix elements of the interface between two different nonlocal homogenized metamaterials.  The scattering matrix elements behave according to boundary conditions similar to Eqs.~(\ref{bound_con_1}-\ref{bound_con_3}).  In this case the discontinuities in the tangential electric and magnetic fields are proportional to the electromagnetic fields on both sides of the interface.  However, the boundary conditions are still determined by the interface parameters in Eqs.~(\ref{bound_con_1}-\ref{bound_con_3}).  That is to say, if one calculates the interface parameters for two different homogenized metamaterials using the scattering model presented in Sec.~\ref{Sec_2}B, those scattering matrix elements can predict the scattering of an interface between those two different nonlocal homogenized metamaterials.

\section{Conclusion}
We have presented a simple model for calculating the scattering matrix for waves scattered from the interface between a homogeneous medium and a homogenized highly symmetric metamaterial crystal.  In doing so we have verified that the relations in Eq.~(\ref{refl_2}) between the various scattering matrix elements are in general not true when spatial dispersion is present.  Using this scattering model, we have postulated a set of phenomenological boundary conditions for the interface of a metamaterial.  This simple model of the metamaterial interface has provided new insights into the non-Maxwellian boundary conditions that exist at the boundaries of metamaterials with spatial dispersion.

\section*{Acknowledgements}
Chris Fietz would like to acknowledge support from the IC Postdoctoral Research Fellowship Program.  Work at Ames Laboratory was supported by the Department of Energy (Basic Energy Science, Division of Materials Sciences and Engineering) under contract no. DE-ACD2-07CH11358.

\appendix
\section{Calculation of s-polarized constitutive parameters}\label{solution}
Ref.~\onlinecite{Fietz_10b} presents a simple and surprisingly successful one dimensional metamaterial homogenization model based on the same principle of replacing layers of a metamaterial crystal with metasurfaces that was used in this paper.  However, the formulas presented in Ref.~\onlinecite{Fietz_10b} are for a p-polarized plane wave ($\textbf{E}=\mathrm{E}_y\hat{\textbf{e}}_y$ and $\textbf{H}=\mathrm{H}_z\hat{\textbf{e}}_z$), while in this paper we consider s-polarized plane waves ($\textbf{E}=\mathrm{E}_z\hat{\textbf{e}}_z$ and $\textbf{H}=\mathrm{H}_y\hat{\textbf{e}}_y$).  The calculation of the constitutive parameters for an s-polarized wave are very similar.  Most of the changes can be made with the electromagnetic duality transformation $\mathrm{E}_z\rightarrow\mathrm{H}_z$, $\mathrm{H}_y\rightarrow-\mathrm{E}_y$, $\epsilon_b\rightarrow\mu_{b}$, and $\mu_b\rightarrow\epsilon_b$~\cite{Jackson_Duality}, though care must be taken with the new definitions of the field amplitude coefficients.  We now briefly derive equations for the s-polarized one dimensional metamaterial homogenization model.

The one dimensional homogenization model is a one dimensional system with all electromagnetic fields harmonic in time with frequency $\omega$.  The one dimensional periodic spatial domain is separated into unit cells of length $a$.  In the center of each cell is a metasurface with surface polarizability $\hat{\alpha}$ defined in Eq.~(\ref{alpha_value}) from the scattering matrix elements of a single layer of the crystal to be homogenized.  The space between each metasurface is filled with a homogeneous material with a background (substrate) permittivity $\epsilon_b$, permeability $\mu_b$ and impedance $z_b=\sqrt{\mu_b/\epsilon_b}$.  For one dimensional s-polarized plane waves, the Maxwellian equations for the microscopic electromagnetic fields in between the metasurfaces are

\begin{equation}\label{Maxwell_eq}
\begin{array}{c}
\displaystyle \frac{\partial\mathrm{e}_z}{\partial x} - \mathrm{i}\frac{\omega}{c} \mu_b\mathrm{h}_y = \frac{\mathrm{I}_y}{c}, \\ \\
\displaystyle \frac{\partial\mathrm{h}_y}{\partial x} - \mathrm{i}\frac{\omega}{c} \epsilon_b\mathrm{e}_z = \frac{\mathrm{J}_z}{c}.
\end{array}
\end{equation}

The metasurfaces interact with each other through free plane waves with frequency $\omega$ and wavenumber $k_0=\sqrt{\epsilon_b\mu_b}\omega/c$.  Across each metasurface, the $\mathrm{e}_z$ and $\mathrm{h}_y$ fields are discontinuous according to Eq.~(\ref{Delta_eq}).  The surface polarization of the metasurfaces which defines the discontinuity is equal to the surface polarizability times the local field strength, as described in Sec.~\ref{Sec_2}.  Finally, the entire system is driven with external electric and magnetic current 

\begin{equation}\label{driving_current}
\mathrm{J}_z = \mathrm{J}_3 e^{\mathrm{i}(\omega t-k_xx)}, \ \ \ \ \ \ 
\mathrm{I}_y = \mathrm{I}_2 e^{\mathrm{i}(\omega t-k_xx)},
\end{equation}

\noindent where the current strengths $\mathrm{J}_3$ and $\mathrm{I}_2$ are chosen arbitrarily.  The solution  to Eq.~(\ref{Maxwell_eq}) with the external current in Eq.~(\ref{driving_current}) is

\begin{equation}
\begin{array}{rl}
\left(\!\!\!\begin{array}{c}
\mathrm{e}_z \\[5pt] \mathrm{h}_y
\end{array}\!\!\!\right) = &
\displaystyle \frac{\mathrm{i} e^{-\mathrm{i}k_xx}}{\epsilon_b\mu_b\omega^2/c^2-k_x^2}
\left(\!\!\!\begin{array}{cc}
\mu_b\omega/c & -k_x \\[5pt] -k_x & \epsilon_b\omega/c
\end{array}\!\!\!\right)\cdot
\frac{1}{c}\left(\!\!\!\begin{array}{cc}
\mathrm{J}_3 \\[5pt] \mathrm{I}_2
\end{array}\!\!\!\right) \\ \\
& + \left(\!\!\!\begin{array}{cc}
a_n e^{-\mathrm{i}k_0 x} + b_n e^{\mathrm{i}k_0 x} \\[5pt]
\displaystyle-\frac{1}{z_b} a_n e^{-\mathrm{i}k_0 x} + \frac{1}{z_b} b_n e^{\mathrm{i}k_0 x}
\end{array}\!\!\!\right).
\end{array}
\end{equation}

\noindent The field coefficients $a_b$ and $b_n$ are the electric field amplitudes of free plane waves propagating in the positive and negative $\hat{\textbf{x}}$-directions  respectively.  These plane waves mediate interactions between adjacent metasurfaces.  The subscript $n$ specifies the metasurface that the plane waves interact with.  For example the $n$ and $n+1$ metasurfaces interact with the right and left moving plane waves with amplitudes $a_n$ and $b_n$ respectively.  These field coefficients are related to each other by the Bloch phase condition

\begin{equation}
\left(\!\!\!\begin{array}{c}
a_{n+1} \\[5pt] b_{n+1}
\end{array}\!\!\!\right) = 
\left(\!\!\!\begin{array}{cc}
e^{-\mathrm{i}(k_x-k_0)a} & 0 \\[5pt]
0 & e^{-\mathrm{i}(k_x+k_0)a}
\end{array}\!\!\!\right)\cdot
\left(\!\!\!\begin{array}{cc}
a_n \\[5pt] b_n
\end{array}\!\!\!\right).
\end{equation}

\noindent The field coefficients are also determined by the boundary conditions across each metasurface given by Eq.~(\ref{Delta_eq}).  The left hand side of Eq.~(\ref{Delta_eq}) is related to the field coefficients by 


\begin{equation}\label{coefficients_1}
\begin{array}{c}
\left(\!\!\!\begin{array}{c}
\Delta \mathrm{h}_z \\[5pt] \Delta \mathrm{e}_y
\end{array}\!\!\!\right) = 
\hat{A}\cdot
\left(\!\!\!\begin{array}{c}
a_b \\[5pt] b_n
\end{array}\!\!\!\right), \\ \\
\hat{A} = \left(\!\!\!\begin{array}{cc}
\!-\displaystyle\frac{1}{z_b} & \displaystyle\frac{1}{z_b} \\[10pt] \ \ 1 & 1
\end{array}\!\!\!\right)\cdot
\left(\!\!\!\begin{array}{cc}
e^{-\mathrm{i}(k_x-k_0)a}-1 & 0 \\[5pt] 0 & e^{-\mathrm{i}(k_x+k_0)a}-1
\end{array}\!\!\!\right).
\end{array}
\end{equation}

\noindent The field strength on the right hand side of Eq.~(\ref{Delta_eq}) is the sum of two solutions, the inhomogeneous solution to Eq.~(\ref{Maxwell_eq}) that does not obey the correct boundary conditions and the homogeneous solutions to Eq.~(\ref{Maxwell_eq}) which corresponds to the free waves mediating the interaction between adjacent metasurfaces.  We choose our coordinates so that $x=0$ at the location of the $n$th metasurface.  The local value of the inhomogeneous solution at the location of the $n$th metasurface is

\begin{equation}\label{coefficients_2}
\begin{array}{c}
\left(\!\!\!\begin{array}{c}
\mathrm{E}_z^{loc} \\[5pt] \mathrm{H}_y^{loc}
\end{array}\!\!\!\right)_{driven} \!\!\!\!\!\!= 
\displaystyle
\mathrm{i}K^{-1}\cdot
\frac{1}{c}\left(\!\!\!\begin{array}{cc}
\mathrm{J}_2 \\[5pt] \mathrm{I}_3
\end{array}\!\!\!\right),
\end{array}
\end{equation}

\noindent where the matrix $\hat{K}$ is defined as

\begin{equation}\label{Maxwell}
\hat{K} \equiv \left(\!\!\!\begin{array}{cc}
\epsilon_b\omega/c & k_x \\[5pt] k_x & \mu_b\omega/c
\end{array}\!\!\!\right).
\end{equation}

The local value of the homogeneous solution at the location of the $n$th metasurface is defined as the average value from both sides of the sheet and is related to the field coefficients by


\begin{equation}\label{coefficients_3}
\begin{array}{c}
\left(\!\!\!\begin{array}{c}
\mathrm{E}_y^{loc} \\[5pt] \mathrm{H}_z^{loc}
\end{array}\!\!\!\right)_{free} \!\!\!\!\!\!= 
\hat{B}\cdot
\left(\!\!\!\begin{array}{c}
a_n \\[5pt] b_n
\end{array}\!\!\!\right), \\ \\
\hat{B} = \left(\!\!\!\begin{array}{cc}
1 & 1 \\[5pt] \displaystyle-\frac{1}{z_b} & \displaystyle\frac{1}{z_b}
\end{array}\!\!\right)\cdot
\displaystyle \frac{1}{2}\left(\!\!\!\begin{array}{cc}
e^{-\mathrm{i}(k_x-k_0)a}+1 & 0 \\[5pt] 0 & e^{-\mathrm{i}(k_x+k_0)a}+1
\end{array}\!\!\!\right).
\end{array}
\end{equation}

\noindent Combining Eq.~(\ref{Delta_eq}) with Eqs.~(\ref{coefficients_1}-\ref{coefficients_3}), we find an equation relating the field coefficients $a_n$ and $b_n$ to the external current.

\begin{equation}\label{ab_driving}
\hat{A}\left(\!\!\!\begin{array}{c} a_n \\[5pt] b_n \end{array}\!\!\!\right) = 
-\displaystyle\frac{\omega}{c}\hat{\alpha} \cdot \hat{K}^{-1} \cdot \frac{1}{c}\left(\!\!\!\begin{array}{c} J_2 \\[5pt] I_3 \end{array}\!\!\!\right)
+\mathrm{i}\frac{\omega}{c}\hat{\alpha} \cdot \hat{B} \cdot \left(\!\!\!\begin{array}{c} a_n \\[5pt] b_n \end{array}\!\!\!\right),
\end{equation}

\noindent which can be solved for the field coefficients to give

\begin{equation}\label{ab_sol}
\left(\!\!\!\begin{array}{c} a_n \\[5pt] b_n \end{array}\!\!\!\right) = \displaystyle \left(\mathrm{i}\frac{\omega}{c} \hat{\alpha} \hat{B}-\hat{A}\right)^{-1}\cdot\frac{\omega}{c}\hat{\alpha}\cdot \hat{K}^{-1}\cdot \frac{1}{c}\left(\!\!\!\begin{array}{c} J_3 \\[5pt] I_2 \end{array}\!\!\!\right).
\end{equation}

\noindent As in Ref.~\onlinecite{Fietz_10b}, the macroscopic polarization density is equal to the polarization of a metasurface divided by the unit cell length $a$.  Using Eq.~(\ref{Delta_eq}), we can relate the field coefficients to the macroscopic polarization yielding
\begin{equation}\label{PJ}
\begin{array}{c}
\left(\!\!\!\begin{array}{c}
\mathrm{P}_z \\[5pt] \mathrm{M}_y
\end{array}\!\!\!\right) = 
\displaystyle\frac{1}{a}
\left(\!\!\!\begin{array}{c}
\mathrm{p}_z \\[5pt] \mathrm{m}_y
\end{array}\!\!\!\right)
=\frac{\hat{\alpha}}{a}\cdot \left[ \mathrm{i}\hat{K}^{-1}\cdot \frac{1}{c}\left(\!\!\!\begin{array}{c} J_3 \\[5pt] I_2 \end{array}\!\!\!\right) + \hat{B}\cdot \left(\!\!\!\begin{array}{c} a_n \\[5pt] b_n \end{array}\!\!\!\right) \right] \\ \\
 = \mathrm{i}\underbrace{\displaystyle\frac{\hat{\alpha}}{a}\left[ 1 - \left(\frac{\omega}{c}\hat{\alpha}+\mathrm{i}\hat{A}\cdot\hat{B}^{-1}\right)^{-1} \cdot\frac{\omega}{c}\hat{\alpha} \right]}_{\hat{Q}}\cdot\hat{K}^{-1}\cdot \displaystyle\frac{1}{c}\left(\!\!\!\begin{array}{c} J_3 \\[5pt] I_2 \end{array}\!\!\!\right).
\end{array}
\end{equation}

\noindent Here we have identified the matrix $\hat{Q}$, defined in Ref.~\onlinecite{Fietz_10b} as the matrix relating the polarization and the external current.  Incidentally, setting the determinant of $\frac{\omega}{c}\hat{\alpha}+\mathrm{i}\hat{A}\cdot\hat{B}^{-1}$ equal to zero is an alternate way of deriving the dispersion relation in Eq.~(\ref{disp_rel}).  Using Eqs.~(\ref{coefficients_1}) and (\ref{coefficients_3}), the matrix $\hat{A}\hat{B}^{-1}$ is evaluated to be

\begin{equation}\label{AB_value}
\hat{A}\hat{B}^{-1} = \displaystyle \frac{-2\mathrm{i}}{\cos(k_xa)+\cos(k_0a)}
\left(\!\!\!\begin{array}{cc}
\sin(k_0a)/z_b & \sin(k_xa) \\[5pt] \sin(k_xa) & z_b\sin(k_0a)
\end{array}\!\!\!\right).
\end{equation}

\noindent Putting this into Eq.~(\ref{PJ}) allows us to evaluate $\hat{Q}$ which we can then use to evaluate the bulk susceptibility $\hat{\chi}$ defined in Ref.~\onlinecite{Fietz_10b} as

\begin{equation}\label{chi_express}
\hat{\chi}(\omega,k_x) = \left(1-\omega/c \hat{Q}\cdot\hat{K}^{-1}\right)^{-1} \cdot\hat{Q}.
\end{equation}

\noindent $\hat{\chi}$ is related to the constitutive tensor by

\begin{equation}\label{constitutive}
\hat{C}(\omega,k_x)=
\left(\!\!\!\begin{array}{cc}
\epsilon_{zz} & \xi_{zy} \\[5pt] \zeta_{yz} & \mu_{yy}
\end{array}\!\!\!\right)=
\left(\!\!\!\begin{array}{cc}
\epsilon_b & 0 \\[5pt] 0 & \mu_b
\end{array}\!\!\!\right)
+\hat{\chi}(\omega,k_x).
\end{equation}

\end{document}